\documentclass[superscriptaddress,prc,twocolumn,showpacs,preprintnumbers,amsmath,amssymb,bibnotes,altaffilletter]{revtex4}
\usepackage{graphicx}
\usepackage{xspace}
\usepackage{sidecap}
\usepackage{amsmath}
\usepackage{amssymb}
\usepackage{url}
\usepackage{braket} 
\usepackage{slantsc}
\usepackage[symbol]{footmisc}
\usepackage[T1]{fontenc}

\def\nuc#1#2{${}^{#1}$#2}

\def\BBz{$\beta\beta(0\nu)$}

\def\BBt{$\beta\beta(2\nu)$}

\def\BB{$\beta\beta$}

\def\Tz{$T^{0\nu}_{1/2}$}

\def\be{\begin{equation}}
\def\ee{\end{equation}}

\begin{document}
\title{The charged current neutrino cross section for solar neutrinos, and background to \BBz\ experiments}


\newcommand{\lanl}{Los Alamos National Laboratory, Los Alamos, NM, USA}
\newcommand{\ou}{Research Center for Nuclear Physics, Osaka University, Ibaraki, Osaka 567-0047, Japan}

\affiliation{\ou} 
\affiliation{\lanl} 

\author{H.~Ejiri}\affiliation{\ou}
\author{S.R.~Elliott}\affiliation{\lanl}

\begin{abstract}
Solar neutrinos can interact with the source isotope in neutrinoless double beta decay experiments through charged current and neutral current interactions. The charged-current product nucleus will then beta decay with a Q-value larger than the double beta decay Q-value. As a result, this process can populate the region of interest and be a background to the double beta decay signal. In this paper we estimate the solar neutrino capture rates on three commonly used double beta decay isotopes, \nuc{76}{Ge}, \nuc{130}{Te}, and \nuc{136}{Xe}. We then use the decay scheme of each product nucleus to estimate the possible background rates in those materials. As half-life sensitivities in future experiments approach $10^{28}$ y, this background will have to be considered.
\end{abstract}
\maketitle

\section{Introduction}
Neutrino-less double beta decay (\BBz) experiments provide a unique opportunity to search for the Majorana neutrino masses, the lepton sector CP phases and other physics quantities beyond the electro-weak standard model. The science motivation for \BBz\ has been well discussed in the literature. In addition many reviews summarize the present status and plans for the future experimental program. References~\cite{ell02, ell04, bar04, eji05, avi05, avi08, bar11, Rode11, Elliott2012, Vergados2012} provide a comprehensive overview of \BBz.

Experiments searching for \BBz\ are reaching ever greater sensitivity. This improvement in sensitivity arises from both an increase in the mass of source isotope and a decrease in background. The next generation of experiments aims to cover the inverted hierarchy region of Majorana neutrino mass (15-50 meV). To meet this goal requires that experiments be capable of measuring half-lives greater than $10^{27}$ y and have backgrounds better than 1 count/(t-y) in the region of interest (ROI) about the \BB\ Q value. Such an experiment would require 1-10 tonnes of isotope. In a future generation of experiments that may try to reach the normal mass hierarchy region ($<$5 meV), a sensitivity greater than $\sim10^{29}$ y will be required. Depending on the \BBz\ nuclear matrix elements and the weak coupling constants, the size of these experiments may reach into the hundreds of tonnes. Experiments at this scale will have to consider background due to solar-$\nu$ charge current (CC) interactions with the \BB\ isotope.  Background due to neutral current (NC) interactions are much less than those due to CC interactions, as will be discussed later.

In Ref.~\cite{ell04} the possibility of solar-$\nu$ CC interactions giving rise to  background for \BBz\ experiments was discussed. In particular the case of \nuc{136}{Xe} was of interest because such background could not be eliminated by the detection of the \nuc{136}{Ba} daughter. However, at the time of that publication, the level structure and transition strengths for \nuc{136}{Cs} were not well-enough known to estimate the size of this background. With the recent measurements of the pertinent nuclear physics input (see Refs.~\cite{eji00a,Vergados2012} for a review), an estimate can now be made. In this article, we calculate the rate for this important isotope and others of frequent use in \BBz\ experiments. Specifically we discuss this CC capture background for \nuc{76}{Ge}, \nuc{130}{Te}, and \nuc{136}{Xe} and discuss possible background rates in general at the \BBz\ ROI. The important case of \nuc{100}{Mo} has been considered in detail elsewhere~\cite{eji00} and so we do not include it here. It should be mentioned, however, that the solar-$\nu$ capture rate is high enough for this isotope that it is also considered for use as a detector for that purpose. The \BB\ isotopes \nuc{116}{Cd}~\cite{Zuber2003} and \nuc{150}{Nd}~\cite{Zuber2011} have been considered as targets for solar neutrino detection. Reference~\cite{Zuber2003} also estimates the \nuc{7}{Be} solar $\nu$ rate for \nuc{130}{Te} and finds a larger value than reported here. The possibility of the elastic scattering of solar neutrinos with electrons in the detector material has been considered previously~\cite{ell04,deBarros2011}.

\section{\BBz\ and Background from Solar $\nu$ Charged Current Interactions}

\begin{figure}[!htbp]
\begin{center}
\includegraphics[width=9cm]{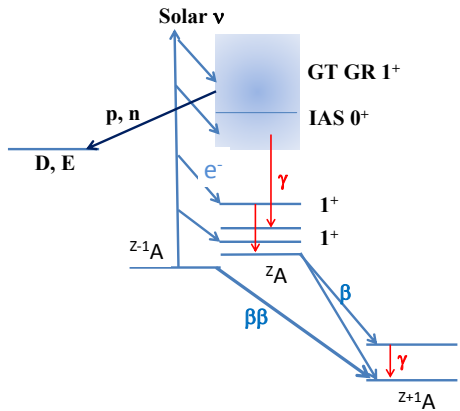}
\caption{A generic energy level diagram showing the double beta decay of an isotope $^{Z-1}A$ to isotope $^{Z+1}A$ with the indicated solar neutrino capture on isotope $^{Z-1}A$ producing isotope $^{Z}A$. The resulting $\gamma$ decay of the excited states in $^{Z}A$ followed by the $\beta$ decay to isotope $^{Z+1}A$ are indicated. The possible particle decays of the excited states in $^{Z}A$ are also shown.}
\label{fig:SolarDBD}
\end{center}
\end{figure}

Figure~\ref{fig:SolarDBD} depicts the various processes under consideration in this manuscript. The \BBz\ of $^{Z-1}A$ to $^{Z+1}A$ shows a peak at the decay Q-value in the \BB\ energy sum spectrum. The intermediate states in $^{Z}A$ are necessarily higher in energy than the initial state $^{Z-1}A$ in most \BB\ nuclei,  and therefore the successive single $\beta$ decays of $^{Z-1}A$ to $^{Z}A$ to $^{Z+1}A$ are energetically forbidden. The solar-$\nu$ CC interaction with the initial nucleus $^{Z-1}A$ excites GT 1$^+$ states and the isobaric analog state 0$^+$  (IAS) in the intermediate nucleus $^{Z}A$, while emitting an electron ($e^-$). If the product nucleus is in an excited state below the particle separation threshold, it will decay to the ground state of $^{Z}A$ emitting some number of $\gamma$ rays. This is followed by the $\beta$ decay of $^{Z}A$ to $^{Z+1}A$. In many cases this $\beta$ decay is to excited states and $\gamma$ rays are emitted also. Alternatively, if the product nucleus is produced in an excited state above the particle separation threshold, it will decay by emitting a proton or neutron to a neighboring nucleus indicated by D and E in the figure.  

The production rate and energy of the particles produced by solar $\nu$ reactions may be comparable to those of the sought \BBz\ signal. Accordingly these processes may contribute as a background within the \BBz\ ROI.

Charge exchange reaction (CER) experiments on \BB\ nuclei have been carried out at RCNP Osaka to study  GT 1$^+$ states involved in \BB\ matrix elements~\cite{eji00,Thies2012,Thies2012a,Thies2012b,Puppe2011,Puppe2012,Guess2011,Frekers2013}. In fact, it was shown that the solar $\nu$ signal could be well measured by a \nuc{100}{Mo} \BB\ experiment on the basis of the observed GT strengths for the solar-$\nu$ CC interaction~\cite{eji00}. In general, there are several 1$^+$ states in the low excitation region and one 0$^+$ IAS and one broad GT 1$^+$ giant resonance (GR) in the high excitation region in the intermediate nucleus. It is these states that are excited by solar-$\nu$ CC interactions with \BB\ nuclei. 

The half-life of $^{Z}A$ for the cases considered here are long enough that the $\nu$ capture and the $\beta$ decay events are separated in time. Therefore we consider three separate processes that may lead to background.

{\bf The solar-$\nu$ CC capture:} will produce $^{Z}A$ in either its ground state or one of a number of excited states.  This capture produces an $e^-$ and, if the nucleus is in an excited state below the particle emission threshold, a number of $\gamma$ rays as it relaxes to its ground state. The sum of the $e^-$ and $\gamma$ ray energies will produce a spectrum that surpasses the \BBz\ Q value and therefore is a potential background.

{\bf The capture to states above the particle emission threshold:} will result in the emission of a $p$ or $n$ and then a number of $\gamma$ rays in the final nucleus. The broad GT 1$^+$ GR and the 0$^+$ IAS are strongly excited by \nuc{8}{B} solar-$\nu$ CC interaction. The upper part of the GT GR and the IAS are above the particle threshold energy and decay by emitting a neutron or proton to low lying states in a neighboring nucleus (indicated by D and E in Fig.~\ref{fig:SolarDBD}). These low lying excited states decay by emitting $\gamma$ rays to the ground state in D or E. The sum of the $e^-$, $p/n$, and $\gamma$ rays also produce a spectrum that may populate the \BBz\ ROI.

{\bf The $\beta$ decay of $^{Z}A$:} will emit a $\beta$ ray and some number of $\gamma$ rays as it decays to the ground state of $^{Z+1}A$. The Q value of this decay is greater than the \BBz\ Q value and therefore it may also populate the \BBz\ ROI.  

All three of these processes have continuum spectra extending beyond the \BBz\ ROI. Consequently, they may contribute to the background at the ROI in future \BB\ experiments with a large mass of isotope. There are also time and spatial relationships between these processes. Hence the contributions depend not only on the nuclear structure of the isotopes, but also the experimental design. Here we summarize some key points.

\begin{itemize}
\item Low lying 1$^+$ states located well below the pp or \nuc{7}{Be} $\nu$ spectral endpoint values of 420 keV and 860 keV are strongly excited by those neutrinos. 
\item The broad GT 1$^+$ giant resonance (GR) and the 0$^+$ IAS are strongly excited by \nuc{8}{B}-$\nu$ CC interactions. The strengths are around $B(GT)\sim1.5(N-Z)$ and $B(F)\sim(N-Z)$ with $N$ and $Z$ being the neutron and proton numbers of the \BB\ nucleus. The lower part of the GR, that is below the particle emission threshold energy, decays by emitting $\gamma$ rays to the ground state of $^{Z}A$, which in turn decays by emitting $\beta$ and $\gamma$ rays to the ground state of $^{Z+1}A$.
\item Since the $e^-$-$\gamma$, $e^-$-$p/n$-$\gamma$ and $\beta$-$\gamma$, sum energy spectra are continua, their contributions to the \BBz\ ROI are proportional to the energy width of the ROI, that is to the experimental energy resolution. 
\item In \nuc{100}{Mo} and \nuc{116}{Cd} nuclei, the intermediate 1$^+$ ground state in $^{Z}A$ is strongly excited by pp and/or \nuc{7}{Be} neutrinos. This state decays predominantly to the ground state of the final nucleus $^{Z+1}A$.  Since the $e^-$ and $\beta$ rays are not followed by $\gamma$ rays, an analysis rejecting events with multiple energy deposits is not possible. If the initial capture $e^-$ and subsequent $\beta$ decay can be correlated, the events may be rejected.  
\item In other \BB\ nuclei, low lying 1$^+$ excited states in the intermediate nucleus $^{Z}A$ are excited by the solar-$\nu$ CC interaction. The ground state in $^{Z}A$ decays to the excited state in the final nucleus $^{Z+1}A$. Consequently $e^-$ and $\beta$ decays are followed by $\gamma$ rays from the excited states in $^{Z}A$ and $^{Z+1}A$, respectively. These $\gamma$ rays can be used to reduce the possible background by a multiple energy deposit analysis~\cite{eji05}.  
\item The $e^-$-$\gamma$ rays produced in the initial interaction producing $^{Z}A$ and the $\beta$-$\gamma$ rays in the final nucleus $^{Z+1}A$ are separated in time by the half life ($\tau_{1/2}$) of $^{Z}A$. Therefore they can be correlated effectively by a single site time correlation analysis (SSTC)~\cite{eji05}. In this way they can be rejected as background. To avoid reduction of the \BBz\ signal efficiency due to accidental coincidences, $\tau_{1/2}$ must be much shorter than the average background interval of 1/$R_b$. Here $R_b$ is the rate of the total background above the detector threshold per volume cell ($V_c$) of the \BB\ detector, and the cell volume is defined in terms of the position resolution of the experiment. In case of a typical high-purity \BB\ detector with background rates below 10 mBq/t, a $V_c$ of 125 mm$^3$, and a density of 4 g/cm$^3$, one finds a maximum possible half-life: $\tau_{1/2}$ $<<$ 0.8 y. Hence SSTC can be applied for most \BB\ detectors with a small $V_c$.
\item The $e^-$ and $\beta$ rays associated with the solar-$\nu$ CC interaction are lone particles from a single point in space but from different points in time, whereas \BBz\ events are two $\beta$ rays from one point in both space and time. Thus measurements of the two $\beta$ tracks are very powerful for selecting the \BBz\ signal and rejecting these solar $\nu$ single $\beta$ rays.
\end{itemize}

\section{Solar Neutrino Capture Rates}

The prescription for calculating the cross section is summarized in Ref.~\cite{eji00a} and we only summarize it here. The cross section ($\sigma_k$) for a neutrino ($\nu$) interacting with a target nucleus ($^{Z-1}A$) to produce the $k^{th}$ excited state in a final nucleus ($^{Z}A_k$),
\begin{equation}
\nu + ^{Z-1}A \rightarrow e^- + ^{Z}A_k,
\end{equation}
is given by the expression
%
\begin{eqnarray}
\label{eqn:CrossSection}
\sigma_k &=& \frac{G^2_Fcos^2\theta_c}{\pi}p_e E_e F(Z,E_e)\left[B(F)_k+(\frac{g_A}{g_V})^2 B(GT)_k\right] \nonumber \\ 
	 &=& (1.597\times10^{-44} \mbox{cm}^2) p_e E_e F(Z,E_e) \nonumber	\\ 
		& & \times\left[B(F)_k+(\frac{g_A}{g_V})^2 B(GT)_k\right]
\end{eqnarray}
%
where $G_F$ is the weak coupling constant, $\theta_c$ is the Cabibo angle, $p_e$ ($E_e$) is the outgoing electron momentum (total energy), $F(Z,E_e)$ is the Fermi function, and $B(F)_k$ ($B(GT)_k$) is the Fermi (Gamow-Teller) response. The constant is applicable for energies and momenta in MeV. $Z$ ($A$) is the atomic number (mass number) of the product. The ratio of the axial vector ($g_A$) and vector ($g_V$) coupling constants is taken to be 1.267~\cite{Mund2013}. We used the tabulated values of $F(Z,E_e)$ from Ref.~\cite{Behrens1969}. The $B(GT)$ values for the states $k$ are derived from the recent series of charge exchange measurements. The $B(F)$ values are taken to be 0, except for the isobaric analog state (IAS) where it is taken to be ($N-Z$).

The solar neutrino reaction rate is then determined by integrating the product of the solar neutrino flux and the cross section given by Eqn.~\ref{eqn:CrossSection} and finally summing over the product nucleus states. Hence, the rate ($R$) is given by
\begin{equation}
R = \sum_k \int \sigma_k\frac{d\phi_{\nu}}{dE_{\nu}} dE_{\nu},
\end{equation}
where $\frac{d\phi_{\nu}}{dE_{\nu}}$ is the neutrino flux as a function of neutrino energy ($E_{\nu}$). We use fluxes from BP05(OP)~\cite{bah05} and express $R$ in units of SNU (10$^{-36}$ interactions per target atom per second).

For the isotopes of interest to this paper, the rates are summarized in Table~\ref{tab:Rates}. We used the charge-exchange-measurement data for \nuc{76}{Ge}~\cite{Thies2012}, \nuc{100}{Mo}~\cite{Thies2012a}, \nuc{130}{Te}~\cite{Puppe2012}, and \nuc{136}{Xe}~\cite{Frekers2013} to obtain the $B(GT)_k$ and $B(F)$ for the IAS for the indicated isotopes. We estimate the uncertainty in our calculations to be less than 10\%, except for that due to the solar \nuc{8}{B} neutrinos, which agrees to about 10-20\%. Given the uncertainties in the $B(GT)$ values, which increase with excited-state energy, this precision suffices.  

The oscillation of solar neutrinos will reduce the CC interaction rate. We approximated that effect by imposing an energy-dependent survival probability ($P_{ee}$) for electron neutrinos exiting the Sun. We parameterized that probability with the following approximate functional form.
%
\begin{equation}
P_{ee} = 0.336+0.117e^{\frac{-E_{\nu}-0.1}{4.82}}+0.119e^{\frac{-E_{\nu}-0.1}{4.88}}
\end{equation}
%
\begin{table}
\caption{The reaction rates in SNU for the various sources of solar neutrinos.}
\label{tab:Rates}
\begin{center}
\begin{tabular}{lcccccc|c}
\hline
Isotope 			& pp & pep & \nuc{7}{Be} & \nuc{8}{B} & \nuc{13}{N} & \nuc{15}{O} & Total \\
\hline\hline
\nuc{76}{Ge}		&0   & 1.4 & 0  			& 13.4 		&  0.1         &    0.8      & 15.6 \\
\nuc{100}{Mo}	&695 &16   &234 			 &16   		&12  			&16  		& 989\\
\nuc{130}{Te}	&0	 & 5.9 &  43.2   	& 15.9   	 & 2.4   		& 4.6  	    & 71.9   \\
\nuc{136}{Xe}	&0   &11.7  &94.8 		&25.8 		&4.3 			&9.1 	   & 145.6\\
\hline
\end{tabular}
\end{center}
\end{table}

\begin{table}
\caption{Same as Table~\ref{tab:Rates} but for oscillated solar neutrino fluxes.}
\label{tab:RatesLMA}
\begin{center}
\begin{tabular}{lcccccc|c}
\hline
Isotope 			& pp & pep & \nuc{7}{Be} & \nuc{8}{B} & \nuc{13}{N} & \nuc{15}{O} & Total \\
\hline\hline
\nuc{76}{Ge}		&0   &0.7  &0    		&5.0 		& 0.06  			&0.4 		 & 6.2  \\
\nuc{100}{Mo}	&390 &8.2  &126  		&6.0   		&6.7  			&8.3  		& 545 \\
\nuc{130}{Te}	&0	 &3.0  &23.2 		&6.1    		&1.3    			&2.4  		& 35.9    \\
\nuc{136}{Xe}	&0   &6.0  &50.9 		&9.8 		&2.3    			 &4.7 		 & 73.4 \\
\hline
\end{tabular}
\end{center}
\end{table}

To compare the solar-$\nu$ CC reaction event rate to that of \BBz, Eqn.~\ref{eqn:BBRate}, taken from Ref.~\cite{bah05}, provides a quick reference. The rate ($R_{\beta\beta}$) of \BBz\ events can be written, assuming 100\%  for the signal selection efficiency,
\begin{equation}
\label{eqn:BBRate}
R_{\beta\beta} = \frac{1}{M}\frac{dN}{dt} = \frac{\lambda N}{M} 
\approx \frac{420}{W}\left(\frac{10^{27}}{T^{0\nu}_{1/2}} 
\right) {\rm /(t-y)}~,
\end{equation}
where the constant is calculated for the molecular weight (W) in g of molecule containing the \BB\ isotope, the \BBz\ half life ($T^{0\nu}_{1/2}$) in y, and the mass ($M$) of the \BBz\ material in tons.

\section{The Initial Solar $\nu$ Capture} 
The initial $\nu$ CC capture emits an electron $e^-$ that will have a continuum of energy up to the \nuc{8}{B} endpoint minus the Q-value for the interaction. The other solar neutrino energies are too low to produce background in the ROI. Here we give a rough estimate of this background so that its impact can be compared to that of the decay of $^{Z}A$.

For \nuc{76}{Ge}, most of the capture is due to the \nuc{8}{B} neutrinos. The energy sum of the $e^-$ and $\gamma$ rays are a continuum spectrum extending up to around 14 MeV. The energy sum spectrum for the $\beta$-$\gamma$ rays from the decay of $^ZA$ extends up to $Q_{\beta} \approx3$ MeV.  Taking into account the reaction Q value, the fraction of the \nuc{8}{B} spectrum that falls within $\Delta E$ window of 4 keV at the \nuc{76}{Ge} \BB\ endpoint is about $3 \times 10^{-4}$. The fraction of the $\beta$-$\gamma$ spectrum from the decay of $^ZA$ that falls in this window is $2 \times 10^{-3}$ (See Section~\ref{Sec:Ge76}.) Hence the ratio is about 15\% indicating that the initial $\nu$ CC capture contributes only a fraction to the background. This estimate is for $^8$B $\nu$ and the case of $^{76}$Ge. The ratio is much smaller in other \BB\ nuclei since other $\nu$ sources are major contributors.    
 (See Section~\ref{Sec:BetaDecay}.)

\section{Solar $\nu$ CC Capture to States above the Particle Separation Energy}
Only the highest energy \nuc{8}{B} neutrinos can populate those states in $^{Z}A$ that are beyond the particle separation threshold. That contribution to the CC interaction rate is a fraction of the total \nuc{8}{B} $\nu$ capture rate. Table~\ref{tab:RatesLMA} gives the capture rates for states below the particle separation energy. If we include that contribution, the \nuc{8}{B} CC rate would increase by 0.9 SNU for \nuc{76}{Ge}, 5.4 SNU for \nuc{100}{Mo} as the IAS is in the middle of the \nuc{8}{B} spectrum. In contrast, it results in small changes for \nuc{130}{Te} and  \nuc{136}{Xe} because the IAS is near the \nuc{8}{B} endpoint. The process will produce a set of particles including $e^-$ and $\gamma$ rays, along with protons or neutrons. The total energy emitted will again be a continuum up to the \nuc{8}{B} endpoint.  Hence this rate will be a subset of the contribution from the initial solar-$\nu$ CC capture rate and is also small compared to the $\beta$ decay of $^{Z}A$.

\section{The $\beta$ Decay of $^{Z}A$}
\label{Sec:BetaDecay}
Of the background processes considered in this paper, this is the most significant and we focus on it for each isotope of interest.

\subsection{Germanium-76}
\label{Sec:Ge76}

A solar neutrino that interacts with \nuc{76}{Ge} through a CC interaction can produce excited states of \nuc{76}{As}. Our estimate of this rate is $\sim$6.2 SNU. Any produced excited state transitions to the \nuc{76}{As} ground state, which decays with a 1.08-d half-life by $\beta$ decay to the ground and excited states in \nuc{76}{Se}. The long half-life of \nuc{76}{As} complicates the use of the solar neutrino event as a tag to remove the As decays on an event-by-event basis but a SSTC might be effective in Ge detectors with a background level less than 100 mBq/t and cell volume $V_c$ = 3 cc.

 Typical Ge \BBz\ experiments have energy resolution near 0.2\% FWHM or $\sim$4 keV. The Q-value for the $\beta$ transition is 2962 keV and when the sum energy of the $\beta$ and $\gamma$ rays fall between 2037 and 2041 keV, it will be a \BBz\ background. We estimated the fraction of decays that will populate this region by considering the three $\beta$ decay transitions with the largest branching-ratios. These three include the 51\% branch to the ground state, the 35.2\% branch to the 559-keV level, and the 7.5\% branch to the 1216-keV level. Hence we considered 93.7\% of the total $\beta$ decay strength and ignored the large number of small branching ratio transitions. This approximation does not significantly alter our result.

\begin{figure}[!htbp]
\begin{center}
\includegraphics[width=9cm]{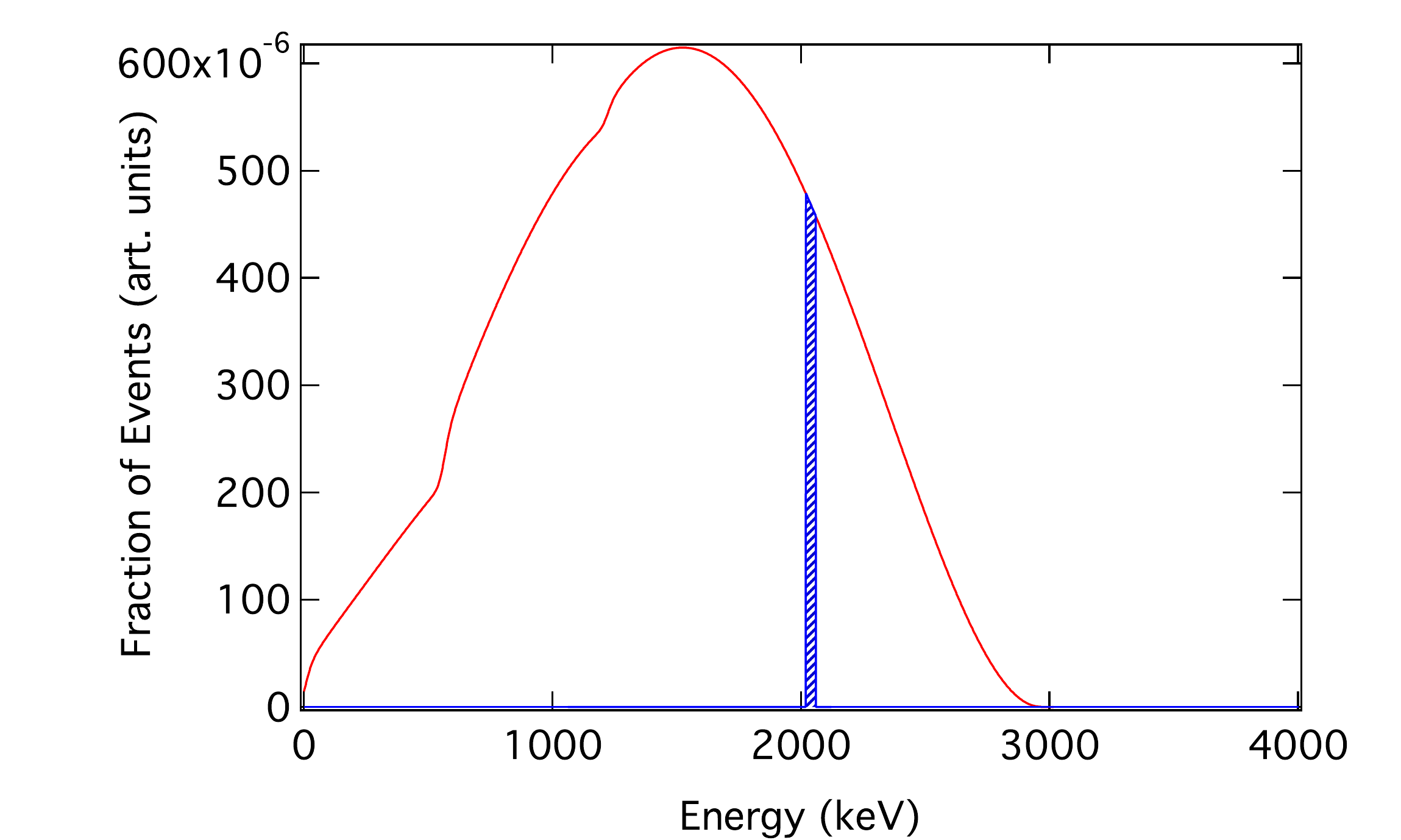}
\caption{The approximate $\beta$-$\gamma$ ray energy spectrum from the decay of \nuc{76}{As}. The shaded region shows the part of the spectrum that will populate the ROI for a resolution of 2\% FWHM.}
\label{fig:As76}
\end{center}
\end{figure}

For the $\beta$ spectrum we used the approximate functional form
\begin{equation}
\frac{dN}{dE} \sim (E_0 - E_e)^2 E_e p_e F(Z,E_e),
\end{equation}
where $E_0$ is the $\beta$ endpoint energy. The calculation indicates that about 0.2\% of the decays will populate the ROI for a resolution of 0.2\% FWHM. Figure~\ref{fig:As76} shows the energy deposited from the $\beta$ and $\gamma$ emissions assuming all the $\gamma$ energy is deposited within the detector. The spectrum has been convolved with a resolution function corresponding to 2\% FWHM. For a resolution of 0.5\% FWHM, the fraction within the ROI is 0.5\% and for 2\% FWHM, it is 1.9\%.

For Ge enriched to 100\% in $^{76}$Ge, Eqn.~\ref{eqn:BBRate} predicts about 5.5
\BBz\ events per year per ton for \Tz\ = 10$^{27}$ y. There are $7.9\times10^{27}$ atoms/t in \nuc{76}{Ge}, therefore we expect about 1.5 solar neutrino events/(t-y) for the 6.2-SNU rate calculated for Ge. Approximately 0.2\% of these, or 0.003 events, will be in the ROI. Different experiments will have different efficiencies for selecting the \BBz\ signal events and for rejecting multiple-sited-energy-deposit $\beta$+$\gamma$ events and therefore removing this background. With no such multiple-site-event (MSE) rejection, the sensitivity floor of a Ge experiment would be about \Tz\ = $1.8\times10^{30}$ y.

\subsection{Telerium-130}

A solar neutrino that interacts with \nuc{130}{Te} through a CC interaction can produce excited states of \nuc{130}{I}. Our estimate of this rate is $\sim$36 SNU. Any produced excited state transitions to the \nuc{130}{I} ground state, which decays with a 12.36-h half-life by $\beta$ decay to excited states in \nuc{130}{Xe}. The long half-life of \nuc{130}{I} complicates the use of the solar neutrino event as a tag to remove the I decays on an event-by-event basis, but a SSTC might be effective in an ultra-pure Te detector with radioactive impurities less than 2 mBq/t within a detector cell around 1 kg.

Because the $\beta$ decay is to a highly excited state that decays quickly emitting accompanying $\gamma$ rays, the fraction of the decays that populate an energy region near the endpoint for the \BBz\ decay of \nuc{130}{I} is relatively large. The various Te \BBz\ experiments have widely varying energy resolutions. The Q-value for the $\beta$ transition is 2949 keV whereas the \BBz\ Q value is 2528 keV. We estimated the fraction of decays that will populate this region by considering the two $\beta$-decay transitions with the largest branching-ratios. These two include the 48\% branch to the 1944-keV level, and the 46.7\% branch to the 2362-keV level. Hence we considered 94.7\% of the total $\beta$ decay strength.

\begin{figure}[!htbp]
\begin{center}
\includegraphics[width=9cm]{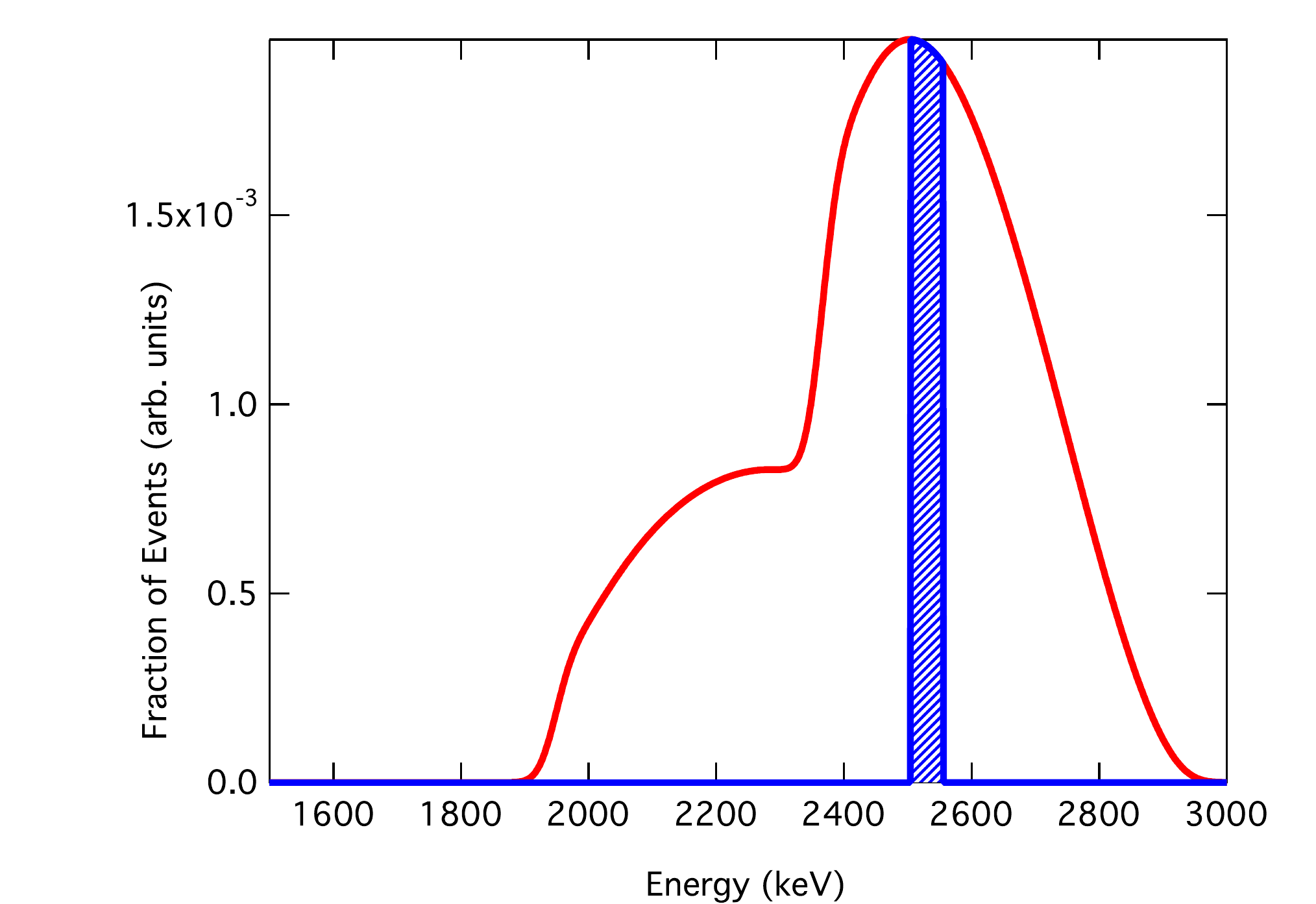}
\caption{The approximate $\beta$-$\gamma$ ray energy spectrum from the decay of \nuc{130}{I}. The shaded region shows the part of the spectrum that will populate the ROI for a resolution of 2\% FWHM.}
\label{fig:I130}
\end{center}
\end{figure}

The $\beta$-spectrum calculation indicates that about 9.8\% of the decays will populate the ROI for a resolution of 2\% FWHM. Figure~\ref{fig:I130} shows the energy deposited from the $\beta$ and $\gamma$ emissions. The spectrum has been convolved with a resolution function corresponding to 2\% FWHM. For a resolution of 0.5\% FWHM, the fraction within the ROI drops to 2.5\% and for 0.2\% FWHM, it is 1\%.

For pure $^{130}$Te, Eqn.~\ref{eqn:BBRate} predicts about 3.2
\BBz\ events per year per ton for \Tz\ = 10$^{27}$ y. There are $4.6\times10^{27}$ atoms/t in \nuc{130}{Te}, therefore we expect about 5.2 solar neutrino events/(t-y) for the 36-SNU rate calculated for Te. Approximately 9.8\% (1\%), or 0.51 (0.05) events for a resolution of 2\% (0.2\%), of these will be in the ROI. Different experiments will have different efficiencies for rejecting MSE-deposit $\beta$+$\gamma$ events and therefore for removing this background. With no such multiple-site-event rejection, the sensitivity floor of a Te experiment would be about  \Tz\ = $6.3\times10^{27}$ y ($6.3\times10^{28}$ y) for the 2\% (0.2\%) resolution.

\subsection{Xenon-136}

A solar neutrino that interacts with \nuc{136}{Xe} through a CC interaction can produce excited states of \nuc{136}{Cs}. Our estimate of this rate is $\sim$74 SNU. Any produced excited state transitions to the \nuc{136}{Cs} ground state, which decays with a 13.16-d half-life by $\beta$ decay to excited states in \nuc{136}{Ba}. The long half-life of \nuc{136}{Cs} prevents the use of the solar neutrino event as a tag to remove the Cs decays on an event-by-event basis, but might permit the implementation of a purification step to remove Cs from the Xe. A SSTC might be effective in liquid Xe detectors with a background level less than 10 mBq/t and $V_c$ = 3 cc. One must consider, however, whether the Cs ion will remain localized for the required time scales within a gaseous or liquid detector medium.

Because the $\beta$ decay is to a highly excited state that decays quickly emitting accompanying $\gamma$ rays, the fraction of the decays that populate an energy region near the endpoint for the \BBz\ decay of \nuc{136}{Xe} is relatively large. Future possible Xe \BBz\ experiments may have an energy resolution near 2\% FWHM or $\sim$50 keV. The Q-value for the $\beta$ transition is 2548 keV and Q-value for \BBz\ is 2458 keV. We estimated the fraction of decays that will populate the ROI by considering the four $\beta$-decay transitions with the largest branching-ratios. These four include the 70.3\% branch to the 2207-keV level, the 13\% branch to the 1867-keV level, the 10.5\% branch to the 2140-keV level, and the 4.7\% branch to the 2053-keV level. Hence we considered 98.5\% of the total $\beta$ decay strength.

\begin{figure}[!htbp]
\begin{center}
\includegraphics[width=9cm]{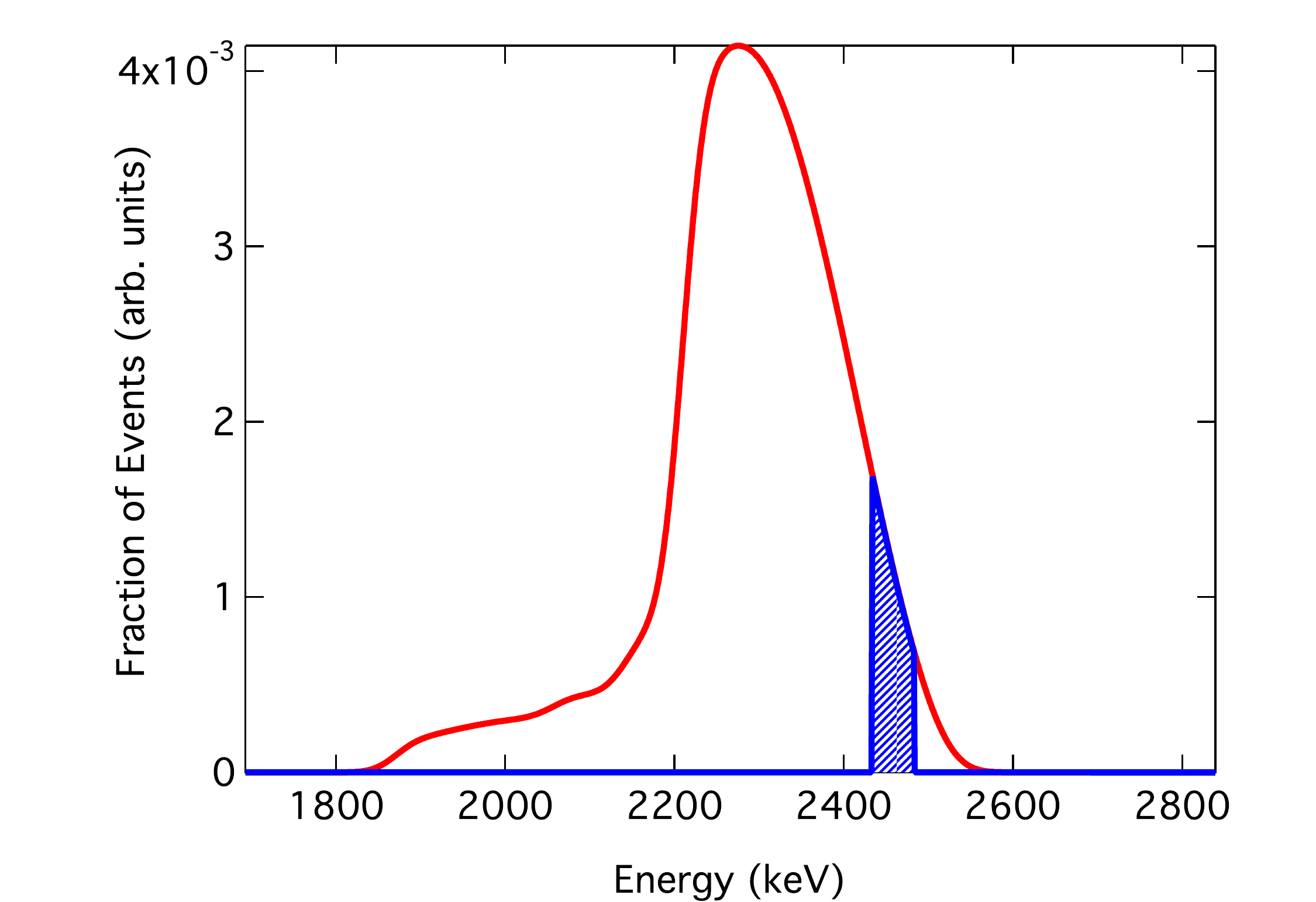}
\caption{The approximate $\beta$-$\gamma$ ray energy spectrum from the decay of \nuc{136}{Cs}. The shaded region shows the part of the spectrum that will populate the ROI for a resolution of 2\% FWHM.}
\label{fig:Cs136}
\end{center}
\end{figure}

The $\beta$-spectrum calculation indicates that about 5.7\% of the decays will populate the ROI for a resolution of 2\% FWHM. Figure~\ref{fig:Cs136} shows the energy deposited from the $\beta$ and $\gamma$ emissions. The spectrum has been convolved with a resolution function corresponding to 2\% FWHM. For a resolution of 0.5\% FWHM, the fraction within the ROI drops to 1.4\% and for 0.2\% FWHM, it is 0.5\%.

For pure $^{136}$Xe, Eqn.~\ref{eqn:BBRate} predicts about 3.1
\BBz\ events per year per ton for \Tz\ = 10$^{27}$ y. There are $4.4\times10^{27}$ atoms/t in \nuc{136}{Xe}, therefore we expect about 10.3 solar neutrino events/(t-y) for the 74-SNU rate calculated for Xe. Approximately 5.7\%, or 0.59 events, of these will be in the ROI. With no MSE rejection, the sensitivity floor of a Xe experiment would be about \Tz\ = $5.3\times10^{27}$ y. It is very important to notice that this background produces a \nuc{136}{Ba} ion and hence will not be rejected by Ba tagging~\cite{moe91}.

\section{Discussion}
Here we considered solar-$\nu$ CC interactions and the subsequent $\beta$-$\gamma$ decays that might contribute to background within the \BBz\ ROI. We have neglected the smaller background contributions from electrons (inverse $\beta$ rays) associated with solar-$\nu$ CC interactions, the solar-$\nu$ CC interactions that excite the product nucleus above proton- or neutron-emission threshold, and solar-$\nu$ NC interactions. Such interactions could add an additional 15\% or more to the total solar-$\nu$ capture rate, depending on the nucleus. We did not consider in detail the various possible proton- or neutron-emission decay channels and how they might produce \BBz\ background. We did not consider NC interactions because their contribution to the ROI is at most 15\% of the major $\beta -\gamma$ BG contribution.   This paper focused on the solar $\nu$ CC interactions with \BB\ isotopes in detectors, but not on interactions with other nuclei. They may not be ignored in detectors where the \BB\ isotopes are only a fraction of the active detector medium. 

We estimated the solar-$\nu$ CC capture rate on \nuc{76}{Ge}, \nuc{130}{Te} and \nuc{136}{Xe} using recently measured Gamow-Teller strengths in charge exchange reactions. We then used those production rates and the known decay schemes of the product nuclei to estimate the potential for background in double beta decay experiments using these isotopes. In the cases of Te and Xe, the product nucleus decays to highly excited states resulting in a sizable fraction of the decays depositing energy in the region of interest for double beta decay. These decays, however, will produce multiple energy deposits by coincident interaction of $\beta$ and $\gamma$ rays, which suggests that position resolution would be effective at reducing this background. We also note that these background $\beta$-$\gamma$ rays from $^Z$A are correlated in time with the electron and $\gamma$ rays from the excited states in $^{Z}A$, and thus they may be reduced by SSTC analyses in case of high purity detectors with good position resolution. Regardless, as experimental half-life sensitivities begin to surpass $5 \times 10^{27}$ y, this background will need to be considered. The background rates within the ROI are approximately proportional to the energy resolution. Table~\ref{tab:Summary} summarizes the results. The energy resolution is indeed very important to reduce  background from the solar-$\nu$ CC interactions as well as those from the \BBt. It is noted, however, that the solar $\nu$ event rate at the ROI may be reduced much by appropriate spatial and time correlation analyses even though one may not avoid the solar-$\nu$ CC interactions with energy resolution as one might \BBt.  

In short, the solar-$\nu$ CC interaction rates are of the same order as anticipated \BBz\ rates in the future \BBz\ experiments searching for an effective Majorana neutrino mass in the inverted to normal hierarchy region ($<$50 meV).  The background event rate due to solar-$\nu$ CC interactions depends strongly on the $B(GT)$ strengths of the low-lying states in the intermediate nuclei as well as the signal selection analysis techniques based on the topology of the event in energy, time and position.  

We note the similar situation occurs also for multi-ton scale dark matter experiments to search for WIMPs in the 10$^{-(10-11)}$ pb region~\cite{Vergados2008}.

\begin{table}
\caption{The background rates in the ROI as a function of resolution given in \% FWHM. These are the raw rates and make no allowance for detector capability to reject events based on time and space correlations.}
\label{tab:Summary}
\begin{center}
\begin{tabular}{|l|cc|}
\hline
	 			& \multicolumn{2}{|c|}{Background Rate}  \\
Isotope 			& \multicolumn{2}{|c|}{(events/(t-y))}  \\
				& 2\%				&		0.2\%						\\
\hline\hline
\nuc{76}{Ge}		& 0.03				&		0.003\\
\nuc{130}{Te}	& 0.51				&		0.05  \\
\nuc{136}{Xe}	& 0.59				&		0.06\\
\hline
\end{tabular}
\end{center}
\end{table}

\section*{Acknowledgments}
We acknowledge support from the Office of Nuclear Physics in the Department of Energy, Office of Science. We gratefully acknowledge the support of the U.S. Department 
of Energy through the LANL/LDRD Program. We thank D. Frekers and R.G.H. Robertson for valuable discussion.

\section{References}
\bibliography{DoubleBetaDecay.bbl}

\end{document}